\documentclass[showpacs,apl,twocolumn,floatfix]{revtex4}

\usepackage{graphicx,float,amsmath}

\begin{document}

\title{Piezoresistance in silicon at uniaxial compressive stresses up to 3 GPa}
\author{J.S. Milne}
\affiliation{Physique de la mati\`ere condens\'ee, Ecole Polytechnique, CNRS, 91128
Palaiseau, France}
\author{I. Favorskiy}
\affiliation{Physique de la mati\`ere condens\'ee, Ecole Polytechnique, CNRS, 91128
Palaiseau, France}
\author{S. Arscott}
\affiliation{Institut d'Electronique, de Micro\'electronique et de Nanotechnologie
(IEMN), CNRS UMR8520, Avenue Poincar\'e, Cit\'e Scientifique, 59652
Villeneuve d'Ascq, France}
\author{Ch. Renner}
\affiliation{Department of Condensed Matter Physics, NCCR Materials with Novel Electronic Properties, University of Geneva, 24 Quai Ernest-Ansermet, CH-1211 Geneva 4, Switzerland}
\author{A.C.H. Rowe}
\email{alistair.rowe@polytechnique.edu}
\affiliation{Physique de la mati\`ere condens\'ee, Ecole Polytechnique, CNRS, 91128
Palaiseau, France}

\begin{abstract}
The room-temperature longitudinal piezoresistance of n-type and p-type crystalline silicon along selected crystal axes is investigated under uniaxial compressive stresses up to 3 GPa. While the conductance ($G$) of n-type silicon eventually saturates at $\approx 45 \%$ of its zero-stress value ($G_0$) in accordance with the charge transfer model, in p-type material $G/G_0$ increases above a predicted limit of $\approx 4.5$ without any significant saturation, even at 3 GPa. Calculation of $G/G_0$ using \textit{ab-initio} density functional theory reveals that neither $G$ nor the mobility, when properly averaged over the hole distribution, saturate at stresses lower than 3 GPa. The lack of saturation has important consequences for strained silicon technologies.
\end{abstract}

\maketitle

In semiconductors such as silicon, applied mechanical stress ($X$) modifies the band structure, principally changing the carrier mobility and hence the conductance, $G$ \cite{smith1954,herring1956,ohmura1990}. This phenomenon is known as piezoresistance. In n-type silicon, uniaxial compressive stress along the $\langle 100 \rangle$ crystal directions lifts the 6-fold degeneracy of the ellipsoidal conduction band minima, resulting in a charge transfer from the four bands perpendicular to $X$ into the two which are parallel with $X$. This charge transfer model \cite{herring1956} dictates an increase in the longitudinal conductivity effective mass ($m^*_{Gl}$) that continues until all the electrons have transfered to the two parallel valleys at which point the resistance change saturates; it correctly describes the piezoresistance of n-type silicon up to several hundred MPa \cite{aubrey1963}. The effect of stress on the valence band of silicon is more complex and cannot be described by a model of equivalent simplicity \cite{adams1954}. Instead, the degeneracy or near-degeneracy of the heavy (HH), light (LH) and split-off (SO) hole states requires the use of simplified analytical \cite{suzuki1984, ohmura1990, kleimann1998} or numerical techniques \cite{richter2008, manku1991, guillaume2006, thompson2006, fan2007} that describe band warping and hole transfer effects. For stresses up to several hundred MPa there is a consensus that the mobility (and hence $G$) change primarily due to a modification of $m^*_{Gl}$, although some authors claim that changes to the inverse scattering rates ($\tau_m$) can be important \cite{sun2007, richter2008}. When using the band edge values of $m^*_{Gl}$, these models are in reasonable agreement with piezoresistance data up to several hundred MPa. In the GPa stress range where there is as yet no experimental data, a saturation of the piezoresistance is predicted once $m^*_{Gl}$ ceases to change with $X$ \cite{thompson2006, sun2007}, somewhat reminiscent of the n-type (and germanium) cases \cite{aubrey1963, cuevas1965}. In contrast, full band Monte Carlo methods \cite{fan2007} predict no such saturation.

Coupled to these fundamental challenges and discrepancies in describing the piezoresistance of p-type silicon is the fact that GPa-range process induced stress has become technologically important since it can be used to reduce $m^*_{Gl}$, thereby improving device speeds and transconductance gains of field-effect transistors \cite{mistry2004, thompson2006, sun2007}. Indeed stressed silicon is one of the key elements of the ITRS roadmap \cite{itrs2009}. The prediction of saturation in the hole mobility (i.e. in the piezoresistance) is thus an important one since it will ultimately limit the increase in transistor performance afforded by strained-silicon technology.

\begin{figure}[!htbp]
\includegraphics[clip,width=8 cm] {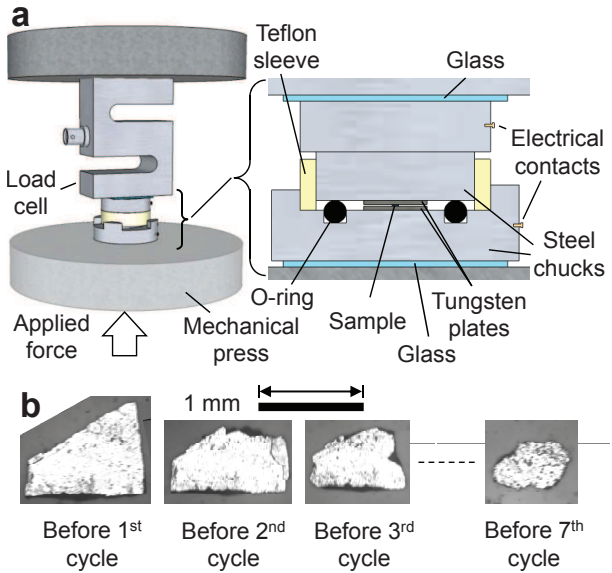}
\caption{(a) Cross-sectional view of the press, chuck assembly and sample, along the with the electrical connections for the conductance measurement. A compressive stress, $X$, is applied parallel to the direction of current flow. (b) A series of photographs of a p-type $\langle 111 \rangle$ sample before each application of stress, showing the cleaving of the sides.}
\label{fig1}
\end{figure}

Here we present piezoresistance data in n- and p-type silicon up to 3 GPa, well above previously experimentally accessible uniaxial stresses, and well beyond the predicted onset of saturation. While the piezoresistance saturation predicted by the charge transfer model holds in n-type silicon up to at least 2.4 GPa, in p-type silicon at even higher stresses no such saturation is observed, in stark contrast to theoretical predictions based on band edge effective masses.

Experimentally, great care must be taken in order to reproducibly apply such large, mechanical stresses. To do so, a mechanical press was adapted to permit two-terminal conductance measurements perpendicularly across bulk semiconducting samples under uniaxial, compressive stress. The assembly depicted in Fig. \ref{fig1}a ensures that the force is applied perpendicularly, by means of a flexible O-ring and a teflon sleeve that maintain parallelicity between the opposing steel chucks. The O-ring, being much softer than the sample, does not affect the applied stress. Flat, mechanically polished tungsten plates are placed between the sample and the chucks to avoid indentation of the steel, as had occurred during initial testing of the assembly. Rigid, metal-covered non-conductors (ceramics and glass) were initially trialled in place of the tungsten but were found to reach their elastic limit and shatter at relatively low stress. The chucks themselves are electrically isolated from the press using glass plates to avoid short circuiting the sample with the press housing. A load cell measures the applied force, which is recorded together with $G$.

The samples were prepared from commercially available double-side-polished n-type [(100), 5-10 $\Omega$.cm, thickness 280 $\mu$m] and p-type [(110), $>$ 1 $\Omega$.cm, thickness 400 $\mu$m] and [(111), 4-7 $\Omega$.cm, thickness 275 $\mu$m] silicon wafers. Low doping densities were chosen to ensure that the measured two-terminal resistance is dominated by the sample itself rather than the leads. Ohmic contacts ($\rho_c = 10^{-6}$ $\Omega $.cm$^2$) are formed by a 100 nm deep thermally activated implant process on both faces of the wafer (doping density $\approx 10^{20}$ cm$^{-3}$) and finished with annealed aluminum contacts. After processing, the samples are cleaved with a diamond tipped scribe into pieces of size $\approx 5\times10^{-7} $m$^{2}$ (see Fig. \ref{fig1}b), so that a pressure of 1 GPa is obtained when about 500 N is applied.

\begin{figure}[!htbp]
\includegraphics[clip,width=8 cm] {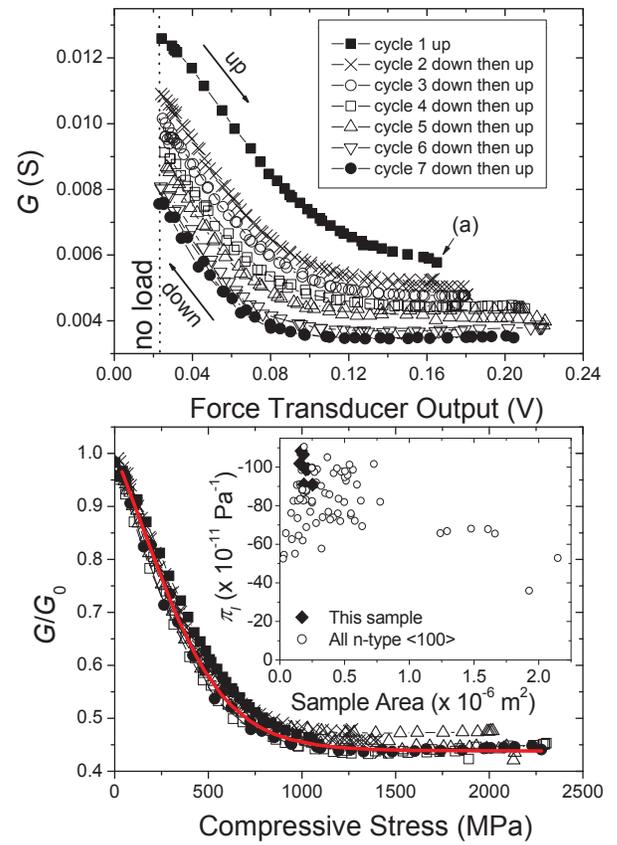}
\caption{(a) Raw data obtained for $X$ parallel to the $\langle 100 \rangle$ crystal direction in n-type silicon, showing discontinuous decreases in conductance due to in-situ cleaving of the sample edges. (b) Data corrected for changes in sample surface area collapse onto a single curve that exhibits saturation above 1 GPa. The fit of the charge transfer model to these curves (solid line) yields $\Xi_u = 12.1$ eV and $\mu_{\perp}/\mu_{\parallel}= 2.9$. The inset shows the measured $\pi_l$ coefficient for this sample (black diamonds) and all other samples (open circles).}
\label{fig2}
\end{figure}

Before measurement, the surface area ($S$) of each sample is estimated from an optical microscope image. In a first cycle, $G$ is measured while the force is gradually applied to the sample until a discontinuous decrease in $G$ is observed (point (a) in Fig. \ref{fig2}a). This jump is accompanied by an audible crack and corresponds to the edges of the sample cleaving off in the press, thereby reducing $S$ and decreasing $G$ (see Fig. \ref{fig1}b). The reduction in $S$ also leads to an instantaneous increase in $X$ since the applied force remains unchanged. In the next cycle, the force is gradually ramped down to zero (without removing the sample) and then slowly ramped up until the sample cracks again. This process is repeated until the entire sample suddenly shatters into a fine powder, presumably because the instantaneous increase in stress resulting from one cracking event causes a cascade of additional cracking events. This typically occurs after about 5 cycles. The resulting raw data is a series of curves relating $G$ to the voltage output of the force transducer, as shown for a (100)-oriented n-type sample in Fig. \ref{fig2}a. These curves cannot be directly compared because $S$ is different for each one, so that the relationship between $X$ and the transducer output is not fixed. Furthermore, the zero-stress conductance ($G_0$) is reduced as $S$ is decreased. $S$ is calculated for each cycle by multiplying the initial value by the ratio of the present and initial values of $G_0$. In this way, the measured force and $S$ can be used to determine $X$ for each cycle. The conductance during each sweep is also normalized to $G_0$, allowing a direct comparison of the piezoresistance for different surface areas. The corrected $G/G_0$ data are plotted against $X$ in Fig. \ref{fig2}b. The curves lie very close to one another and saturation occurs at $\approx$ 1 GPa.

\begin{figure}[!htbp]
\includegraphics[clip,width=8 cm] {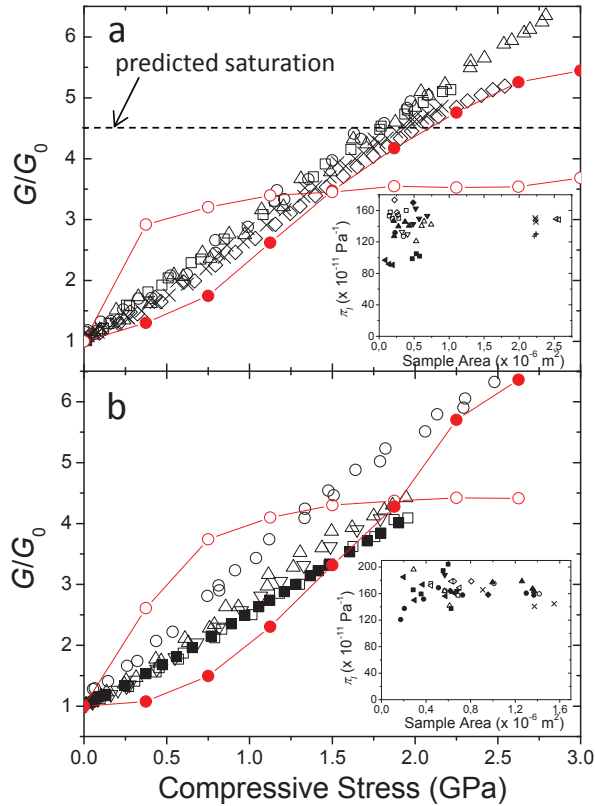}
\caption{Corrected data obtained along the (a) $\langle 110 \rangle$ and (b) $\langle 111 \rangle$ directions in p-type silicon (black symbols). No saturation is observed in either case. The insets show the measured $\pi_l$ coefficients for all samples. In both cases the result of the \textit{ab-initio} calculation is shown when concentration weighted band edge $m_{Gl}^*$ for the HH and LH bands are used (lines with open circles), and when Eq. \ref{K0} is used to average over the LH and HH hole distributions (lines with closed circles).}
\label{fig3}
\end{figure}

The data in Fig. \ref{fig2}b is fitted using Aubrey's adaptation \cite{aubrey1963} of the charge transfer model \cite{herring1956} which has two free parameters; $\Xi_u$, the shear deformation potential and $\mu_{\perp}/\mu_{\parallel}$, the mobility ratio of the ellipsoidal valleys perpendicular and parallel to $X$. The solid curve in Fig. \ref{fig2}b corresponds to this fit and, over all the n-type samples tested, $\Xi_u = 12.6 \pm 3.5$ eV and $\mu_{\perp}/\mu_{\parallel} = 2.6 \pm 0.4$ is found. These values correspond well with those calculated and measured elsewhere \cite{smith1954,herring1956,aubrey1963}, indicating that the experimental method and the conductance correction technique described above are reasonable. Furthermore, the slopes of the $G/G_0$ curves at low stress yield the longitudinal piezoresistance coefficient, $\pi_l$, from which it is found that $\pi_l = \partial G/\partial X \times 1/{G_0} = -82 \pm 15 \times 10^{-11}$ Pa$^{-1}$ which is slightly lower than the generally accepted value \cite{smith1954}. While the difference is not excessive, especially when compared to the know experimental scatter in reported values for the [100] crystal direction \cite{beaty2002}, it may be the result of an unintentionally applied transverse stress. Given the small aspect ratio of the sample it is conceivable that transverse stress cannot relax over the sample thickness. Since the transverse piezoresistance coefficients have the opposite sign to $\pi_l$, this may systematically result in slightly lower apparent $\pi_l$.

Figure \ref{fig3}a and b show the data for p-type silicon with $X \parallel \langle 110 \rangle$ and $ X \parallel \langle 111 \rangle$ crystal directions respectively. For $X \ll 1$ GPa the slopes of these curves yield $\pi_l = 140 \pm 21 \times 10^{-11}$ Pa$^{-1}$ for the $\langle 110 \rangle$ direction and $\pi_l = 159 \pm 37 \times 10^{-11}$ Pa$^{-1}$ for the $\langle 111 \rangle$ direction (see insets in Fig. \ref{fig3}a and Fig. \ref{fig3}b) which is a factor of two higher than the generally accepted low-stress values \cite{smith1954}. In attempting to understand this difference it is important to note that in Smith's original work $X$ is a \textit{tensile} stress between 1 and 10 MPa i.e. smaller than the stress increment in the GPa-adapted apparatus described here. Moreover there is some experimental \cite{shifren2004, tsang2008} and theoretical \cite{guillaume2006, fan2007} evidence in the literature that the p-type piezoresistance along the [110] direction is larger for compressive stresses than for tensile stresses. One experimental observation reports values of $\pi_l$ under 600 MPa of compressive stress that are 1.5 times larger than Smith's value \cite{shifren2004}. This is in excellent agreement with a theory \cite{fan2007} which predicts increases of a factor of 2 in $\pi_l$ at higher stresses i.e. similar to the values measured here. The intermediate stress ($X < 600$ MPa) values of $\pi_l$ reported here (at least for the [110] crystal direction) should probably not therefore be compared directly with Smith's values, but rather with those of Shifren \cite{shifren2004} and Tsang \cite{tsang2008}. The remaining difference may be the result of a lack of parallelicity in the tungsten contacts with the sample surfaces. In this case $X$ will only be applied locally to parts of the sample, and will be larger than its averaged value. This will therefore tend to systematically underestimate $X$ and overestimate $\pi_l$. A final (and most important) aspect of the data is the lack of saturation at high stress for both crystal orientations (see dotted horizontal line in Fig. \ref{fig3}a for the predicted saturation level, \cite{thompson2006, sun2007}). At $X \approx 3$ GPa, $G/G_0 \approx 6.5$, well above the predicted limit of about 4.5. If $X$ is systematically underestimated due to contact parallelicity problems then this result is even more surprising since $X$ is \textit{at least} 3 GPa.

In order to investigate the discrepancy between the predicted saturation in $G/G_0$ and the experimental observations, \textit{ab-initio} density functional calculations of the full Brillouin zone band structure were performed up to 3 GPa (oriented along the $\langle 100 \rangle$, $\langle 110 \rangle$ and $\langle 111 \rangle$ crystal directions) using ABINIT \cite{gonze2009} with the PAW method to account for spin-orbit coupling \cite{torrent2008}. The calculation is performed by imposing a target stress tensor and then performing a structural optimization which yields the deformed lattice parameters. These are then used to calculate the band structure and the density of states. The tetrahedron integration method is modified in order to obtain the partial density of states for the $i^{th}$ band, $D_i(E)$. In the conduction band with $X \parallel <100>$, $G/G_0$ is in excellent agreement with the charge transfer model shown in Fig. \ref{fig2}b (\textit{ab-initio} result not shown). 

In the following discussion of the valence band with $X \parallel <110>$ or $X \parallel <111>$, the total hole concentration is assumed to be a stress-independent $p = 10^{16}$ cm$^{-3}$ which is imposed using the calculated $D_i(E)$ and fixing the Fermi energy as necessary. The GW correction \cite{bruneval06} is used to confirm that the intrinsic concentration change due to the applied stress is indeed negligible compared to $p$. It is also noted that since the semiconductor is in the non-degenerate limit, the calculated values of $G/G_0$ depend only very weakly on the exact value of $p$.

For $X \parallel <110>$ the band edge calculation of $m_{Gl}^*$ for the HH and LH bands, weighted by the respective concentrations, results in a saturation of $G/G_0$ at $\approx 3.5$ (see lines with open circle symbols in Fig. \ref{fig3}a) in qualitative agreement with the behavior predicted elsewhere \cite{sun2007} (see dotted line in Fig. \ref{fig3}a). This clearly does not describe the experimental data. In order to better do so, the calculated dispersion relations for the two bands which contribute significantly to the transport (HH and LH) are used to find the velocity as a function of energy, $v_i(E) \propto \partial E/\partial k|_i$ for the $i^{th}$ band. The partial conductance of the $i^{th}$ band is then proportional to $K_{0i}$, the usual zeroth order moment of the solution to the Boltzmann transport equation \cite{ziman60}: \begin{equation} \label{K0} K_{0i} \propto \int{v_i^2(E) \tau_{mi}(E) \frac{\partial f^0}{\partial E} D_i(E) \mathrm{d}E}. \end{equation} Here $\partial f^0/\partial E$ is the Fermi window function with $f^0$ the equilibrium Fermi-Dirac distribution function. In order to properly evaluate Eq. \ref{K0}, the energy dependence of the relaxation time, $\tau_{mi}(E)$, must be calculated. Here an oversimplified but useful approach is taken: it is assumed that momentum relaxation is determined by intraband optical phonon scattering. Given the angular dependence of the interband and intraband overlap factors for the scattering rates \cite{jacoboni83}, the error induced by this approximation is probably not too large. In this case \cite{ziman60} $\tau_{mi}(E) \propto 1/D_i(E)$ and $K_{0i}$ no longer depends on $D_i(E)$. $G$ is then calculated according to $G \propto \sum_i K_{i0}$. The resulting dependence of $G/G_0$ on $X$ (shown in Fig. \ref{fig3}a, lines with solid circles) exhibits no saturation and is surprisingly close to the measured data. It is noted that a full band Monte Carlo simulation \cite{fan2007} yields similar results that are in excellent agreement with the experimental data presented here.

\begin{figure}[!htbp]
\includegraphics[clip,width=8 cm] {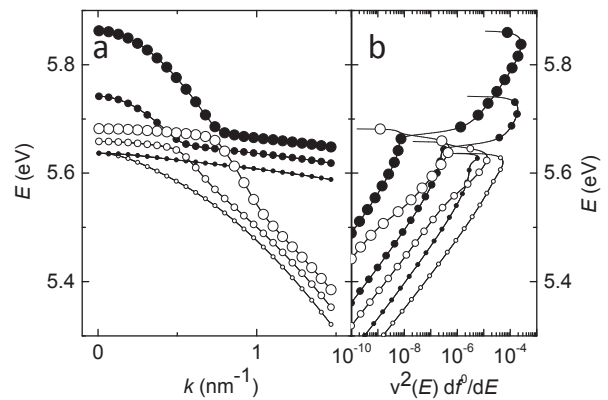}
\caption{(a) The LH (open circles) and HH (closed circles) band structure calculated along $\langle 110 \rangle$ parallel to $X$ for three stress levels indicated by the circle size: 0 (small), 1.5 (medium) and 3 GPa (large). LH/HH anti-crossing plays an important role in determining ensemble averaged $G$ for $X < 1$ GPa where the integrand of Eq. \ref{K0} is non-negligible (b). At high stresses, holes progressively occupy only the HH resulting in a gradual roll-off in $G$ without hard saturation.}
\label{fig4}
\end{figure}

Fig. \ref{fig4}a	details the calculated changes in band structure oriented along a $\langle 110 \rangle$ direction, parallel to $X$. In this case the LH (HH) band is represented by open (closed) circles for three values of $X$: 0, 1.5 and 3 GPa. Stress increases the energy of the HH band maximum at $k = 0$ and results in a LH/HH anti-crossing for $k \neq 0$. For $X < 1$ GPa, the anti-crossing is close to the zone center and significantly modifies the ensemble averaged effective transport mass (and thus $v(E)$) for both bands. This is important for the overall value of $G$ since the integrand of Eq. \ref{K0} (see Fig. \ref{fig4}b) is non-negligible close to the anti-crossing. At higher stresses the HH/LH splitting increases and the anti-crossing moves out to larger $k$ where the integrand becomes smaller due to the finite size of the Fermi window function. This reflects the physical fact that as $X$ increases, holes are progressively restricted to the HH band at energies well above the anti-crossing. It is also the reason for the lack of hard saturation in $G/G_0$ since the thermal distribution of holes does not abruptly fall to zero close to the band edge, but is rather smeared out over the band dispersion. In this very simple description a saturation in $G/G_0$ is still expected as can be seen in the slight roll-off in the calculated curve in Fig. \ref{fig3}a, although hard saturation will presumably only occur for $X \gg 3$ GPa (note also that indeed a weak onset of saturation is sometimes observed as in Fig. \ref{fig3}a (open diamonds)). 

For $X \parallel \langle 111 \rangle$ the behaviour is qualitatively similar to the $\langle 110 \rangle$ direction. A saturation in $G/G_0$ is predicted (see lines with open circles in Fig. \ref{fig3}b) when accounting only for the concentration weighted (HH and LH) band edge values of $m_{Gl}^*$ whereas the ensemble average shows no saturation whatsoever. Again, the application of Eq. \ref{K0} yields a surprisingly good agreement with the experimental data (see lines with solid circles in Fig. \ref{fig3}b).

The concrete experimental evidence presented here along with a simple theoretical interpretation strongly suggest that up to 3 GPa at least, the variation of $G/G_0$ with $X$ is due entirely to the ensemble averaged mobility variation, and not to a change in $p$. Moreover, it also suggests the mobility variation arises from \textit{both} a change in $m^*_{Gl}$ due to charge transfer and band warping, \textit{and} to an increase in $\tau_m$ with compressive stress. The qualitative agreement between experiment and theory tentatively indicates that the interband scattering contribution to $\tau_m$ is negligible. These conclusions have potentially important consequences for p-type strained-silicon devices, particularly in the technologically important $\langle 110 \rangle$ directions. They indicate that much larger hole mobilities than previously envisaged are possible at process-induced stresses higher that those which are currently accessible in p-type transistor channels, suggesting that further investment in this technology may yield dividends.

\acknowledgements{This work was supported by the ANR (PIGE ANR-2010-021). The authors thank F. Bruneval for advice on the use of ABINIT (abinit.org), B. Xu for useful discussions concerning the calculation of $\tau_m$, and T. Verdier and T. Porteboeuf for their technical help.}

\bibliographystyle{apsrev}
\bibliography{achr}

\end{document}